\documentclass[runningheads,fleqn]{svmult}
\usepackage{makeidx}   
\usepackage{graphicx}  
\usepackage{subeqnar}  
\usepackage{multicol}  
\usepackage{taphys}    
\makeindex             
\usepackage{amsfonts}
\usepackage{amssymb}
\usepackage{amsbsy}
\newcommand{\wit}[1]{\widetilde{#1}}    

\begin{document}
\title*{Local scale-invariance in ageing phenomena}
\toctitle{Local scale-invariance in ageing phenomena}
\titlerunning{Local scale-invariance \& ageing}
\author{Malte Henkel}
\authorrunning{Malte Henkel}
\institute{Laboratoire de Physique des Mat\'eriaux (CNRS UMR 7556), 
           Universit\'e Henri Poincar\'e Nancy I,
           B.P. 239, F -- 54506 Vand{\oe}uvre l\`es Nancy Cedex,
           France}

\maketitle              

\begin{abstract}
Many materials quenched into their ordered phase undergo ageing and there show
dynamical scaling. For any given dynamical exponent $z$, this can be extended
to a new form of local scale-invariance which acts as a dynamical symmetry.
The scaling functions of the two-time correlation and response functions of
ferromagnets with a non-conserved order parameter are determined. These
results are in agreement with analytical and numerical studies of various
models, especially the kinetic Glauber-Ising model in 2 and 3 dimensions.
\end{abstract}

\noindent 
Ageing\index{ageing} in its most general sense refers 
to the change of material properties as a function of time. 
In particular, {\em physical ageing}\index{phyical ageing} occurs when
the underlying microscopic processes are reversible while on the other hand,
biological systems age because of irreversible chemical reactions going on
within them. Historically, ageing phenomena were first observed in 
glassy systems, see \cite{Stru78},
but it is of interest to study them in systems without disorder. These should 
be conceptually simpler and therefore allow for a better understanding.
Insights gained this way may become useful for a later 
study of glassy systems.

\section{Phenomenology of ageing}
%
In describing the phenomenology of ageing system, we shall refer throughout 
to simple ferromagnets, see \cite{Bray94,Bouc00,Godr02,Cugl02} for
reviews. We consider systems which undergo a second-order equilibrium 
phase transition at a critical temperature $T_c>0$ and we shall assume
throughout that the dynamics admits no macroscopic conservation law. Initially,
the system is prepared in some initial state (typically one considers an
initial temperature $T_{\rm ini}=\infty$).
The system is brought out of equilibrium by quenching it to a final 
temperature $T\leq T_c$. Then $T$ is fixed and the system's temporal
evolution is studied. It turns out that the relaxation back to global 
equilibrium is very slow (e.g. algebraic in time) with a formally infinite 
relaxation time for {\em all} $T\leq T_c$. 

Let $\phi(t,\vec{r})$ denote the time- and space-dependent order parameter 
and consider the two-time correlation\index{correlation function} 
and (linear) response functions\index{response function}
\begin{equation}
C(t,s;\vec{r}) = \langle \phi(t,\vec{r})\phi(s,\vec{0})\rangle \;\; , \;\;
R(t,s;\vec{r}) = \left.
\frac{\delta\langle\phi(t,\vec{r})\rangle}{\delta h(s,\vec{0})}
\right|_{h=0}
\end{equation}
where $h$ is the magnetic field conjugate to $\phi$ and space-translation
invariance was already implicitly assumed. 
The autocorrelation\index{autocorrelation function} and 
autoreponse functions\index{autoresponse function} are given by
$C(t,s) = C(t,s;\vec{0})$ and $R(t,s) = R(t,s;\vec{0})$ 
where $t$ is referred to as {\em observation time}\index{observation time} 
and $s$ is called the {\em waiting time}\index{waiting time}. 

\begin{figure}
\includegraphics[width=.45\textwidth]{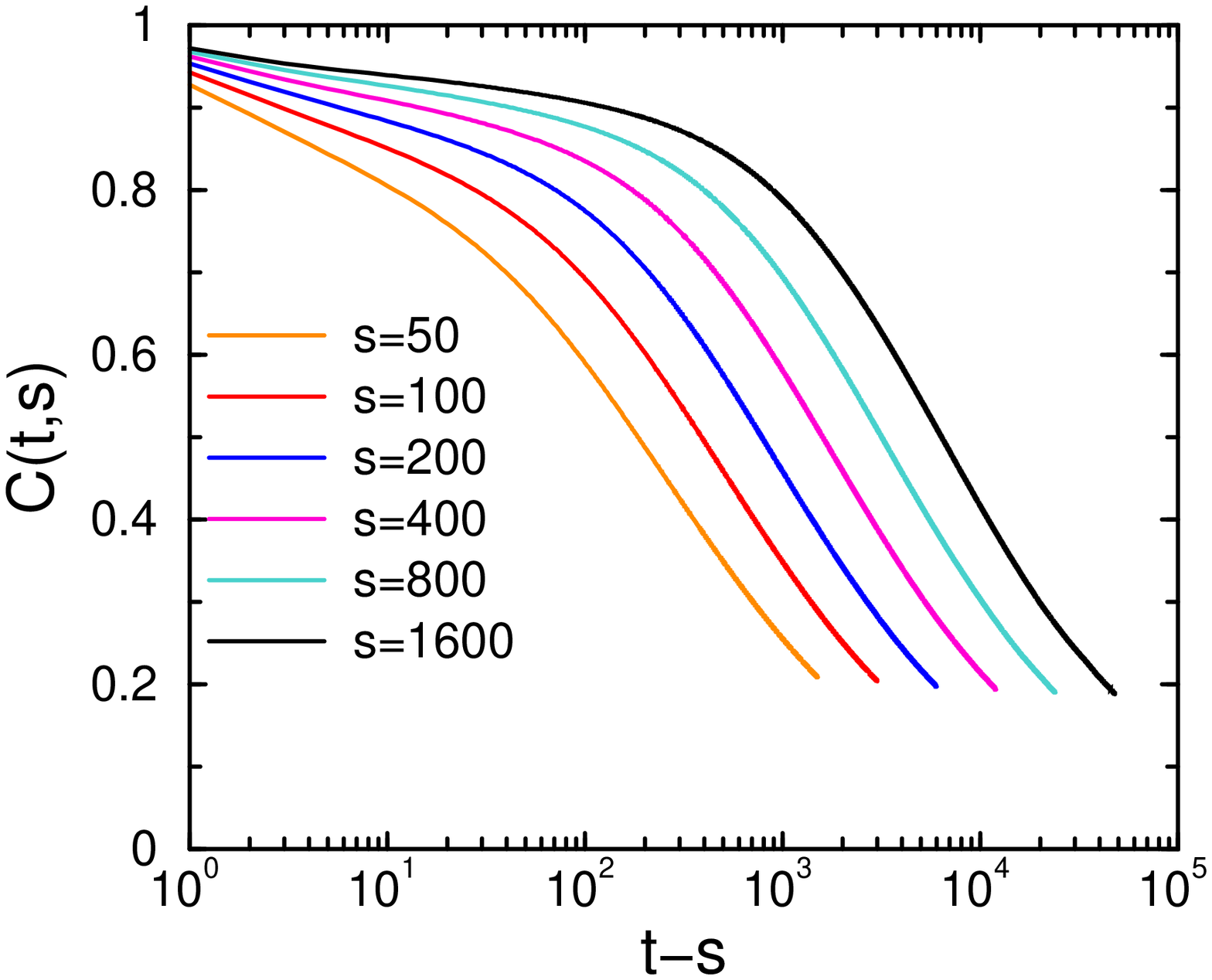}
\includegraphics[width=.45\textwidth]{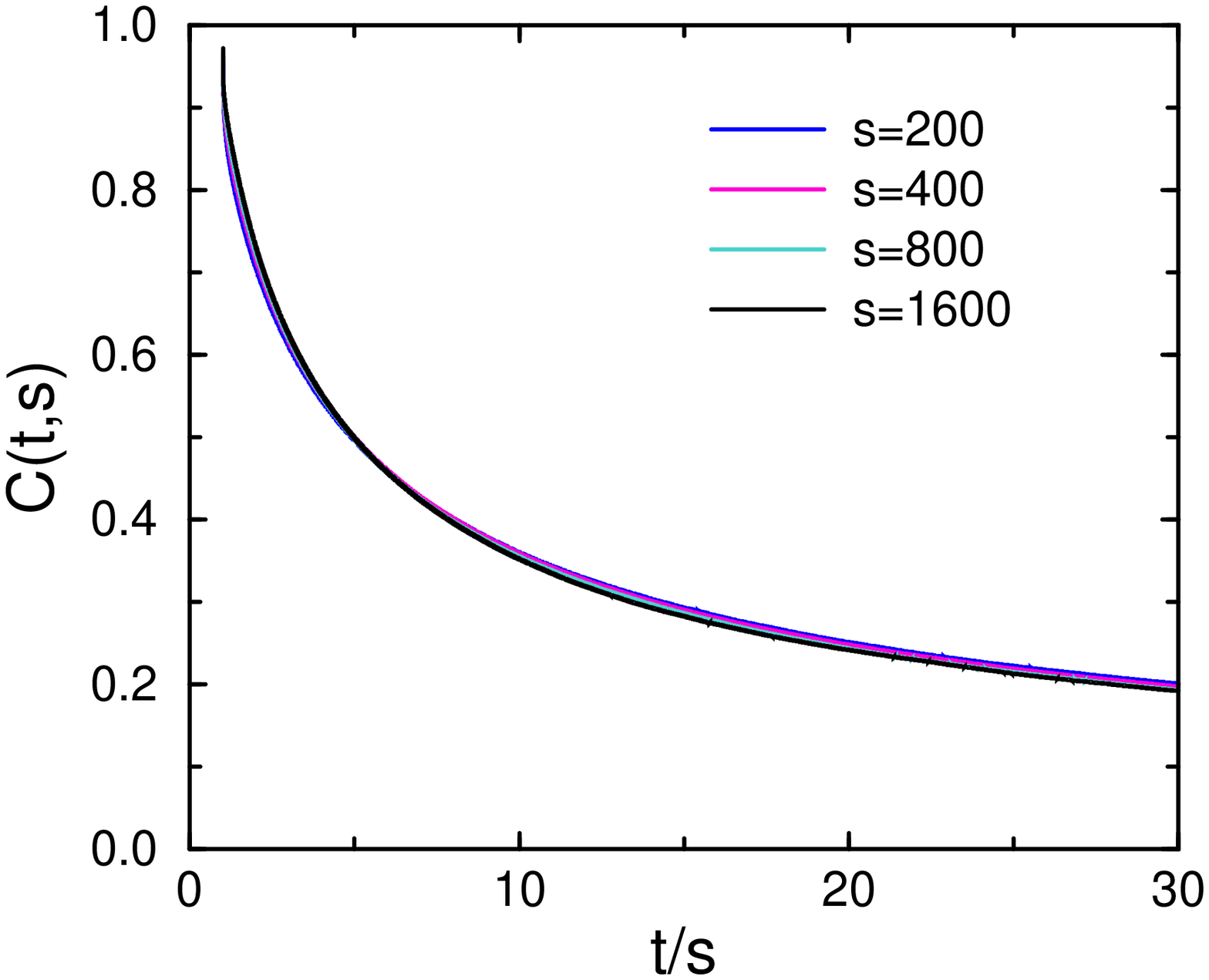}
\caption[]{Two-time autocorrelator\index{autocorrelation function} 
in the $2D$ Glauber-Ising model,\index{Glauber-Ising model} with temperature 
$T=1.5=0.66 T_c$ on a $600^2$ lattice and a disordered initial state.}
\label{Abb1}
\end{figure}

In figure~\ref{Abb1} we show the autocorrelator $C(t,s)$ of
the $2D$ kinetic Ising model with Glauber dynamics after a quench 
to the final temperature $T=1.5$. 
In the left panel,
the dependence of $C(t,s)$ on the time difference $\tau=t-s$ is shown. 
Clearly, the autocorrelator depends on both $t$ and $s$. For large values of
$s$ and $\tau\lesssim s$, the values of $C(t,s)$ reach a quasistationary value 
$C_{\rm qs}(\tau)\simeq M_{\rm eq}^2$, where $M_{\rm eq}$ is the equilibrium
magnetization. In the regime $\tau\gtrsim s$ one observes an algebraic decay
of $C(t,s)$. Qualitatively similar
behaviour is known from glassy systems and the simultaneous dependence of
$C(t,s)$ and/or $R(t,s)$ on {\em both} $t$ and $s$ is the formal definition
of {\em ageing behaviour}\index{ageing, definition}. The strong dependence of
$C(t,s)$ on the waiting time (which expresses the sensibility of the system's
properties on its entire history) seems at first sight to lead to 
irreproducible data and hence to prevent a theoretical understanding
of the ageing phenomenon. Remarkably, Struik \cite{Stru78} observed in polymeric
glasses submitted to mechanical stress that the linear responses of quite
distinct materials could be mapped onto a single and {\em universal}
master curve. We illustrate this here in the ferromagnetic Glauber-Ising model 
through the data collapse in the right panel of figure~\ref{Abb1}. 
Remarkably, a dynamical scaling\index{dynamical scaling} holds although
the equilibrium state need not be scale-invariant. 

On a more microscopic level, correlated domains of a linear size $L(t)$ form.
These are ordered if $T<T_c$ but do contain internal long-range 
fluctuations at criticality. In the first case, the system undergoes 
{\em phase-ordering kinetics}\index{phase-ordering kinetics} and in the second 
{\em non-equilibrium critical dynamics}\index{non-equilibrium critical 
dynamics}. For sufficiently large times, the domain size scales with time as
\begin{equation}
L(t) \sim t^{1/z}
\end{equation}
where $z$ is the dynamical exponent\index{dynamical exponent}. The slow 
relaxation to global equilibrium (although local equilibrium is rapidly
achieved) comes about since for $T< T_c$ there are at least two distinct and
competing equilibrium states. These states merge at $T=T_c$. On each site 
$\vec{r}$ the local environment selects the local equilibrium state. 

As suggested from figure~\ref{Abb1}, one expects a scaling 
regime\index{ageing regime} to occur when
\begin{equation} \label{1:ageReg}
t\gg \tau_{\rm micro} \;\; , \;\; 
s\gg \tau_{\rm micro} \;\; , \;\; 
t-s\gg \tau_{\rm micro}
\end{equation}
where $\tau_{\rm micro}$ is some `microscopic' time scale. We shall see 
later how important the third condition in (\ref{1:ageReg}) is. 
If the conditions (\ref{1:ageReg}) hold, one expects \cite{Bray94,Godr02}
\begin{subeqnarray}
C(t,s) &=& s^{-b} f_C(t/s) \;\;\;\;\;\;, \;\; 
f_C(y)\sim y^{-\lambda_C/z} \;\; ; \;\; y\to\infty \label{1:C}
\\
R(t,s) &=& s^{-1-a} f_R(t/s) \;\; , \;\; 
f_R(y)\sim y^{-\lambda_R/z} \;\; ; \;\; y\to\infty \label{1:R}
\end{subeqnarray}
These scaling forms should hold for both $T<T_c$ and $T=T_c$ although
the values of the exponents will in general be different in these two cases. 
Here $\lambda_C$ and $\lambda_R$ are the 
autocorrelation\index{autocorrelation exponent} \cite{Fish88}
and autoresponse\index{autoresponse exponent} \cite{Pico02} exponents, 
respectively. They are independent of the equilibrium exponents and 
of $z$ \cite{Jans89}. 
It was taken for granted since a long time that 
$\lambda_C=\lambda_R$ but examples to the contrary have recently been found for
spatial long-range correlations in the initial data \cite{Pico02} and 
in the random-phase sine-Gordon model \cite{Sche03}. If $T_{\rm ini}=\infty$,
the inequality $\lambda_C=\lambda_R\geq d/2$ holds \cite{Yeun96}. 

\begin{table}
\caption{Values of the non-equilibrium exponents $a$, $b$ and $z$ 
for non-conserved ferromagnets with $T_c>0$. The non-trivial critical-point 
value $z_c$ is model-dependent.\label{Tab1}}
\renewcommand{\arraystretch}{1.4}
\setlength\tabcolsep{10pt}
\begin{tabular}{lllll}  
\hline\noalign{\smallskip}
        & $a$            & $b$            & $z$   & Class \\ 
\noalign{\smallskip}\hline\noalign{\smallskip}
$T=T_c$ & $(d-2+\eta)/z$ & $(d-2+\eta)/z$ & $z_c$ & L \\ \hline
$T<T_c$ & $(d-2+\eta)/z$ & 0              & 2     & L \\
        & $1/z$          & 0              & 2     & S \\ \hline
\end{tabular}
\end{table}

The values of the non-equilibrium exponents\index{non-equilibrium exponents} 
$a$ and $b$ apparently depend on properties
of the equilibrium system as follows \cite{Henk02a} and are listed in 
table~\ref{Tab1}, together with those of $z$. 
We restrict to non-conserved ferromagnetic systems with $T_c>0$. 
If the equilibrium order parameter correlator 
$C_{\rm eq}(\vec{r})\sim \exp(-|\vec{r}|/\xi)$ with
a finite $\xi$, the system is said to be of {\em class S} and if 
$C_{\rm eq}(\vec{r})\sim |\vec{r}|^{-(d-2+\eta)}$, 
it is said to be of {\em class L}. At criticality, a system is always in 
class L, but if $T<T_c$, systems such
as the Glauber-Ising model are in class S, whereas the kinetic spherical model
is in class L. For class S, the value of $a$ comes from the well-accepted
idea \cite{Bert99,Bouc00} that the time-dependence of macroscopic averages 
comes from the motion of the domain walls. For class L, it follows from a
hyperscaling argument \cite{Henk03e}. 

Having fixed the values of the critical exponents, we can state our main 
question: what can be said on the form of the universal scaling functions 
$f_C(y), f_R(y)$ in a general, model-independent way~?

\section{Local scale-invariance}
%
Our starting point is the rich evidence, accumulated through many decades
and reviewed in \cite{Bray94}, in favour of dynamical scale-invariance in 
ageing phenomena. 
The order parameter field $\phi=\phi(t,\vec{r})$ scales
\begin{equation} \label{2:cov}
\phi(t,\vec{r}) = \mathfrak{b}^{-x_{\phi}} \phi( \mathfrak{b}^{-z} t,
\mathfrak{b}^{-1} \vec{r})
\end{equation}
where $\mathfrak{b}$ is a constant rescaling factor. We now ask whether 
eq.~(\ref{2:cov}) can be sensibly generalized to general space-time
dependent rescalings $\mathfrak{b}=\mathfrak{b}(t,\vec{r})$ 
\cite{Henk94,Henk97,Henk02}. This ansatz can be motivated as follows. 

\noindent 
{\bf Example 1}. \index{conformal invariance} 
Consider an equilibrium critical point in $(1+1)$-dimensional 
space-time. Then $z=1$ and let $w=t+{\rm i} r$ and $\bar{w}=t-{\rm i} r$. Any
angle-perserving space-time transformation is conformal and is given by the
analytic transformations $w\mapsto f(w), \bar{w}\mapsto \bar{f}(\bar{w})$. 
A well-known result from field theory states \cite{Card96} 
that for short-ranged interactions, there is a 
Ward identity\index{Ward identity} such that invariance under space- and 
time-translations, rotations and dilatations implies 
conformal invariance. Furthermore, basic quantities as the order parameter 
are {\em primary} under the conformal group and transform as 
$\phi(w,\bar{w})\mapsto 
(f'(w)\bar{f}'(\bar{w}))^{x_{\phi}/2}\phi(f(w)\bar{f}(\bar{w}))$ \cite{Bela84}. 
Hence $n$-point correlation functions and the values of the exponents
$x_{\phi}$ can be found exactly from conformal symmetry, see e.g.
\cite{Card96,Henk99} for introductions. Here we merely need
the projective conformal transformations 
$f(w)=(\alpha w+\beta)/(\gamma w+\delta)$ 
with $\alpha\delta-\beta\gamma=1$. Fields which transform covariantly
under those are called {\em quasiprimary} \cite{Bela84}. The associated 
infinitesimal transformations are 
$\ell_n=-w^{n+1}\partial_w$, $n\in\{\pm 1,0\}$, and 
satisfy the Lie algebra $[\ell_n,\ell_{n'}]=(n-n')\ell_{n+n'}$. 

\noindent 
{\bf Example 2}. \index{Schr\"odinger-invariance} 
Let $z=2$ and consider $d$ space dimensions. 
The {\em Schr\"odinger group}\index{Schr\"odinger group} 
{\sl Sch}($d$) is defined by \cite{Nied72}
\begin{equation}
t\mapsto \frac{\alpha t+\beta}{\gamma t +\delta} \;\; , \;\;
\vec{r}\mapsto \frac{{\cal R}\vec{r}+\vec{v}t+\vec{a}}{\gamma t +\delta} 
\;\; ; \;\;
\alpha\delta-\beta\gamma=1
\end{equation}
where ${\cal R}\in SO(d)$, $\vec{a},\vec{v}\in\mathbb{R}^d$ and 
$\alpha,\beta,\gamma,\delta\in\mathbb{R}$. It is well-known that {\sl Sch}($d$) 
is the maximal kinematic group of 
the free Schr\"odinger equation ${\cal S}\psi=0$ with 
${\cal S}=2m{\rm i}\partial_t - \partial_{\vec{r}}^2$ \cite{Nied72} 
(that is, it maps any solution of ${\cal S}\psi=0$ to another solution). 
There are many Schr\"odinger-invariant systems, e.g. non-relativistic 
free fields \cite{Hage72} or the Euler equations of 
fluid dynamics \cite{ORai01}. As in the conformal case,
for local theories there is a Ward identity\index{Ward identity}  
such that \cite{Henk03}
\begin{equation} \label{2:GalInv}
\left. \begin{array}{l}
\mbox{\rm space translation invariance} \\
\mbox{\rm scale invariance with $z=2$} \\
\mbox{\rm Galilei invariance} 
\end{array} \right\} \Longrightarrow
\mbox{\rm Schr\"odinger invariance}
\end{equation}
We point out that Galilei invariance\index{Galilei-invariance} 
has to be required and for applications
to ageing we note that time-translation invariance is not needed. Indeed, 
a non-trivial Galilei-invariance is only possible for a complex wave 
function $\psi$. In applications to ageing, we shall identify below 
the `complex conjugate' of the order parameter $\phi^*=\wit{\phi}$ with the 
response field of non-equilibrium field theory \cite{Henk03}. 
We denote the 
Lie algebra of {\sl Sch}($d$) by $\mathfrak{sch}_d$. Specifically,
$\mathfrak{sch}_1=\overline{\{X_{\pm 1,0}, Y_{\pm 1/2}, M_0\}}$ with the 
non-vanishing commutation relations
\begin{eqnarray}
\left[ X_{n}, X_{n'}\right] &=& (n-n') X_{n+n'} \;\; , \;\;
\left[ X_{n}, Y_{m}\right] = \left(\frac{n}{2}-m\right) Y_{n+m} \;\; , \;\;
\nonumber \\
\left[ Y_{1/2}, Y_{-1/2}\right] &=& M_0
\end{eqnarray}
where $n,n'\in\{\pm 1, 0\}$ and $m\in\{\pm 1/2\}$.

\noindent 
{\bf Example 3}. 
For a dynamical exponent $z\ne 2$, we construct infinitesimal
generators of local scale transformations from the following requirements
\cite{Henk02} (for simplicity, set $d=1$): (a) Transformations in time 
are $t\mapsto(\alpha t+\beta)/(\gamma t+\delta)$ with 
$\alpha\delta-\beta\gamma=1$.
(b) The generator for time-translations is $X_{-1}=-\partial_t$ and for 
dilatations $X_0=-t\partial_t-z^{-1}r\partial_r-x/z$,
where $x$ is the scaling dimension of the fields $\phi,\wit{\phi}$ on which
the generators act. (c) Space-translation invariance is required, with
generator $-\partial_r$. Starting from these conditions, we can show by
explicit construction that there exist generators $X_n$, $n\in\{\pm 1, 0\}$, 
and $Y_m$, $m=-1/z, 1-1/z,\ldots$ such that
\begin{equation} \label{Henkel:gl2-10}
\left[ X_{n}, X_{n'}\right] = (n-n') X_{n+n'} \;\; , \;\;
\left[ X_{n}, Y_{m}\right] = \left(\frac{n}{z}-m\right) Y_{n+m} 
\end{equation}
For generic values of $z$, it is sufficient to specify the `special' 
generator \cite{Henk02}
\begin{equation} \label{Henkel:gl2-11}
X_1 = -t^2\partial_t -Ntr\partial_r -Nxt - \wit{\alpha}r^2\partial_t^{N-1}
-\wit{\beta} r^2\partial_r^{2(N-1)/N} -\wit{\gamma}\partial_r^{2(N-1)/N} r^2
\end{equation}
from which all other generators can be recovered and where we wrote $z=2/N$ 
and $\wit{\alpha},\wit{\beta},\wit{\gamma}$ 
are free constants (further non-generic solutions exist for $N=1$ and $N=2$). 
For $z=2$ we recover the Schr\"odinger Lie algebra $\mathfrak{sch}_1$. 
Now, the condition $[X_1, Y_{N/2}]=0$ is only satisfied if either 
(I) $\wit{\beta}=\wit{\gamma}=0$ which we call {\em type I} or else 
(II) $\wit{\alpha}=0$ which we call {\em type II} \cite{Henk02}. 

\noindent {\bf Definition:} If a system is invariant under the generators of
either type I or type II it is said to be 
{\em locally scale-invariant}\index{local scale-invariance} of
type I or type II, respectively. 

Local scale-invariance of type I can be used to describe strongly
anisotropic equilibrium critical points. The application to Lifshitz points in 
$3D$ magnets with competing interactions is discussed in \cite{Plei01,Henk02}. 
The generators of type II are suitable for applications to ageing
phenomena and will be studied here. First, we note that the generators
$X_n, Y_m$ form a kinematic symmetry of the linear differential equation 
${\cal S}\psi=0$ where ${\cal S}=-z^2(\wit{\beta}+\wit{\gamma})\partial_t +
\partial_{r}^z$ \cite{Henk02}. Recently, systems of non-linear equations
invariant under these generators with $\wit{\alpha}=\wit{\beta}=\wit{\gamma}=0$
but extended to an infinite-dimensional symmetry $t\mapsto f(t)$ 
have been found \cite{Cher04}. Second, we consider the consequences for the
scaling form of the response function $R(t,s;\vec{r})$. To do this, we
recall that in the context of Martin-Siggia-Rose theory (see \cite{Jans92}) a
response function $R(t,s)=\langle\phi(t)\wit{\phi}(s)\rangle$ may be viewed as
a correlator. If both $\phi$ and $\wit{\phi}$ transform as quasiprimaries, the
hypothesis of covariance of the autoresponse function leads to the two
conditions $X_0 R = X_1 R=0$. Of course, ageing systems cannot be invariant
under time-translations. From the explicit form of the generators given above
these equations are easily solved and the result can be compared with the
expected asymptotic behaviour (\ref{1:R}). This leads to the general
result\index{response function}
\cite{Henk01,Henk02}
\begin{equation} \label{2:autoR}
R(t,s) =r_0 \Theta(t-s)\left(\frac{t}{s}\right)^{1+a-\lambda_R/z} 
\left( t-s\right)^{-1-a}
\end{equation}
where $r_0$ is a normalization constant and the causality condition $t>s$ is
explicitly included. Furthermore, the space-time response is given by
$R(t,s;\vec{r})=R(t,s)\Phi\left(r (t-s)^{-1/z}\right)$ where $\Phi(u)$ 
solves the equation \cite{Henk02}
\begin{equation}
\left[\partial_u +z\left(\wit{\beta}+\wit{\gamma}\right)u\partial_u^{2-z}
+2z(2-z)\wit{\gamma}\partial_u^{1-z}\right] \Phi(u)=0
\end{equation}
In the special case $z=2$, this reduces to \cite{Henk94}
\begin{equation} \label{2:Rtsr}
R(t,s;\vec{r}) = R(t,s)\exp\left(-\frac{\cal M}{2}\frac{\vec{r}^2}{t-s}\right)
\end{equation}
where ${\cal M}=\wit{\beta}+\wit{\gamma}$ is constant. 

We point out that the derivation of the space-time response needs the 
assumption of Galilei-invariance (suitably generalized if $z\ne 2$). In turn,
the confirmation of the form (\ref{2:Rtsr}) is a given system undergoing ageing 
provides evidence in favour of Galilei-invariance in that system. We shall next
describe tests of (\ref{2:autoR},\ref{2:Rtsr}) in the Glauber-Ising model in
$d\geq 2$ dimensions before we return to a fuller discussion of
the physical origins of local scale-invariance. 

\section{Numerical test in the Glauber-Ising model}
%
We wish to test the predictions (\ref{2:autoR},\ref{2:Rtsr}) of local 
scale-invariance in the kinetic Glauber-Ising model, defined by the Hamiltonian
${\cal H}=-\sum_{(\vec{i},\vec{j})} \sigma_{\vec{i}}\sigma_{\vec{j}}$ where
$\sigma_{\vec{i}}=\pm 1$. Based on a master equation, we use the heat-bath 
stochastic rule\index{Glauber-Ising model} 
\begin{equation}
\sigma_{\vec{i}}(t+1) = \pm 1 \mbox{\rm ~~with probability~}
\frac{1}{2}\left[ 1\pm\tanh(h_{\vec{i}}(t)/T)\right]
\end{equation}
with the local field $h_{\vec{i}}(t)=\sum_{\vec{n}(\vec{i})}\sigma_{\vec{n}}(t)$
and $\vec{n}(\vec{i})$ runs over the nearest neighbours of the site $\vec{i}$. 

The response function is too noisy to be measured directly, therefore following
\cite{Barr98} one may add a quenched spatially random magnetic field 
$\pm h_{(0)}$ between the times $t_1$ and $t_2$ and measure the integrated
response $M(t,t_1,t_2):=h_{(0)} \int_{t_1}^{t_2} \!\D u\, R(t,u)$. Two schemes
a widely used, namely the `zero-field-cooling' (ZFC) scheme, where $t_1=s$ and
$t_2=t$ and the `thermoremanent' (TRM) scheme, where $t_1=0$ and $t_2=s$. 
However, in both schemes it is not possible to na\"{\i}vely use the scaling 
form (\ref{1:R}) and integrate in order to obtain $M$. This comes about since
in both cases some of the conditions (\ref{1:ageReg}) for the validity of this
scaling form are violated. Taking this fact into account leads to the 
following results \cite{Henk02a,Henk03e}: 
(a) the thermoremanent magnetization\index{thermoremanent magnetization}
\begin{eqnarray}
\rho(t,s) &:=& \int_{0}^{s} \!\D u\, R(t,u) =  r_0 s^{-a} f_M(t/s) 
+ r_1 s^{-\lambda_R/z} g_M(t/s) \label{3:rho} \\
f_M(y) &=& y^{-\lambda_R/z} {_2F_1}\left(1+a,\frac{\lambda_R}{z}-a;
\frac{\lambda_R}{z}-a+1;\frac{1}{y}\right) \; , \;\;
g_M(y)\simeq y^{-\lambda_R/z} \nonumber
\end{eqnarray}
where $r_{0,1}$ are normalization constants. The first term
is as expected from na\"{\i}ve scaling. In practice, 
$a$ and $\lambda_R/z$ are often quite close and the size of the correction term
may well be notable for $T<T_c$ 
(at $T=T_c$, $a$ and $\lambda_R/z$ are usually quite distinct); 
(b) the zero-field-cooled susceptibility\index{zero-field-cooled susceptibility}
\begin{equation}
\chi(t,s) := \int_{s}^{t} \!\D u\, R(t,u) = \chi_0 + s^{-A} g(t/s) +
{\rm O}\left( s^{-a}\right)
\end{equation}
with a constant $\chi_0$ and some scaling function $g$. 
For systems of class S, we have
$A=a-\kappa$, where $\kappa$ measures the width $w(t)\sim t^{\kappa}$ of the
domain walls \cite{Henk03e}. In the Glauber-Ising model, one has $\kappa=1/4$ 
in $2D$ and $w(t)\sim \sqrt{\ln t\,}$ in $3D$ \cite{Abra89}, 
while $\kappa=0$ for $d>3$. 
Consequently, the term of order $s^{-a}$ coming from na\"{\i}ve scaling 
is not even the dominant one in the long-time limit $s\to\infty$ and a simple 
phenomenological analysis of data of $\chi(t,s)$ is likely to produce 
misleading results. For systems of class L, $A=0$. 

Indeed, based on high-quality numerical MC data for $\chi(t,s)$ in the 
$2D$ Glauber-Ising model and performing a straightforward scaling analysis 
according to $\chi(t,s)\sim s^{-a}$ but {\em without} taking 
the third condition (\ref{1:ageReg}) for the validity of scaling into account, 
it had been claimed that 
$a=1/4$ in that model \cite{Corb03}. However, that
analysis is based on the identification $A\stackrel{?}{=}a$ which cannot be
maintained. Rather, for the $2D$ Glauber-Ising model, one has $a=1/2$ and
$\kappa=1/4$, reproducing $A=1/4$ in agreement with the MC data.

\begin{table}
\caption{Values of the autoresponse exponent $\lambda_R$ 
and of the parameters $r_0$, $r_1$ and $\cal M$ in the 
Glauber-Ising model for an infinite-temperature initial state. \label{tab2}}
\renewcommand{\arraystretch}{1.4}
\setlength\tabcolsep{10pt}
\begin{tabular}{llllll}  \hline\noalign{\smallskip}
$d$ & $T$ & $\lambda_R$ & \multicolumn{1}{c}{$r_0$} & 
\multicolumn{1}{c}{$r_1$} & \multicolumn{1}{c}{$\cal M$} \\ 
\noalign{\smallskip}\hline\noalign{\smallskip}
2   & 1.5 & 1.26  & $~~~1.76 \pm 0.03$  & $-1.84 \pm 0.03$   & $4.08\pm 0.04$ \\
3   & 3   & 1.60  & $~~~0.10 \pm 0.01$  & $~~\,0.20\pm 0.01$ & $4.22\pm 0.05$ \\
\hline
\end{tabular}
\end{table}
 
\begin{figure}
\includegraphics[width=.6\textwidth]{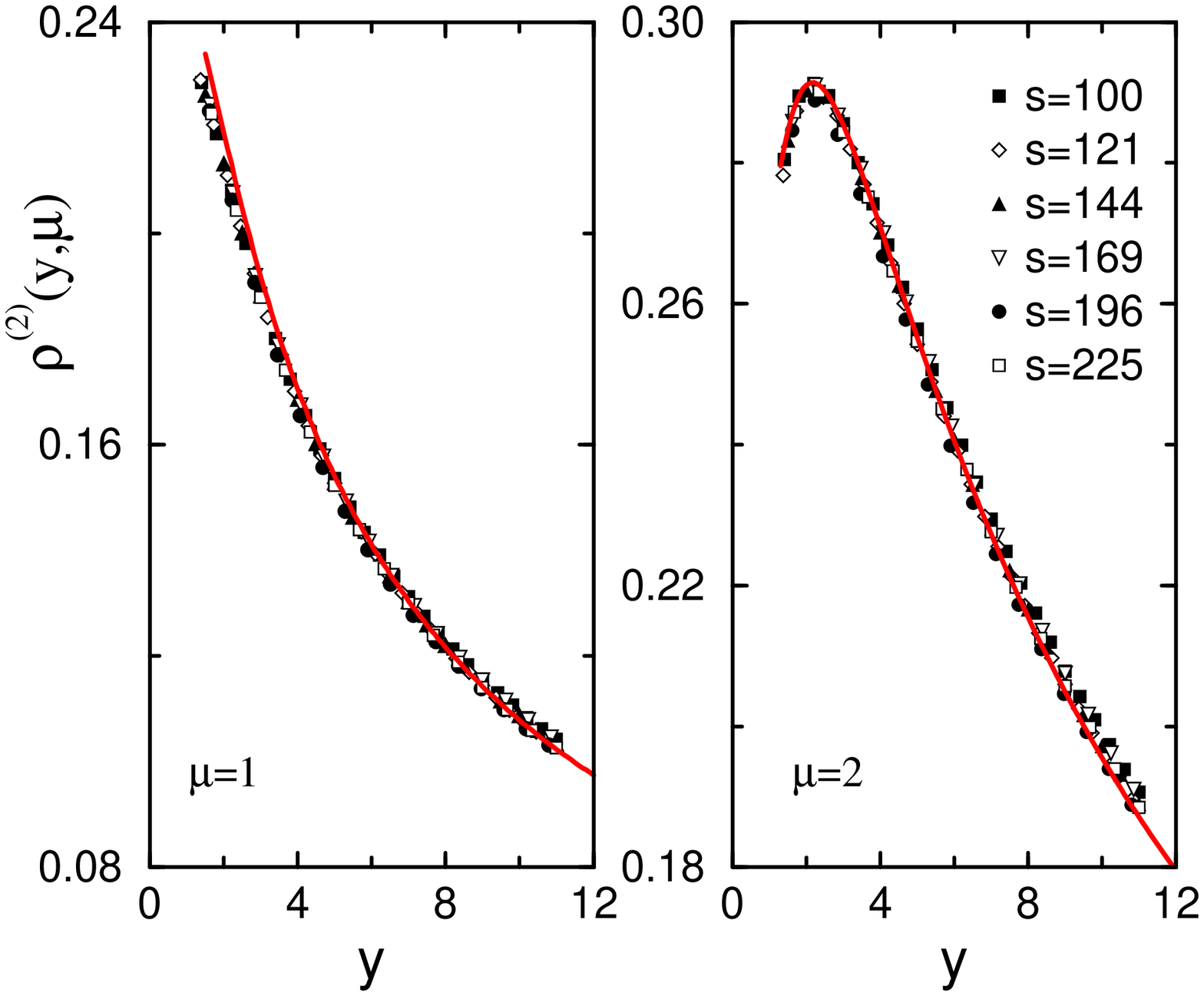}
\includegraphics[width=.3\textwidth]{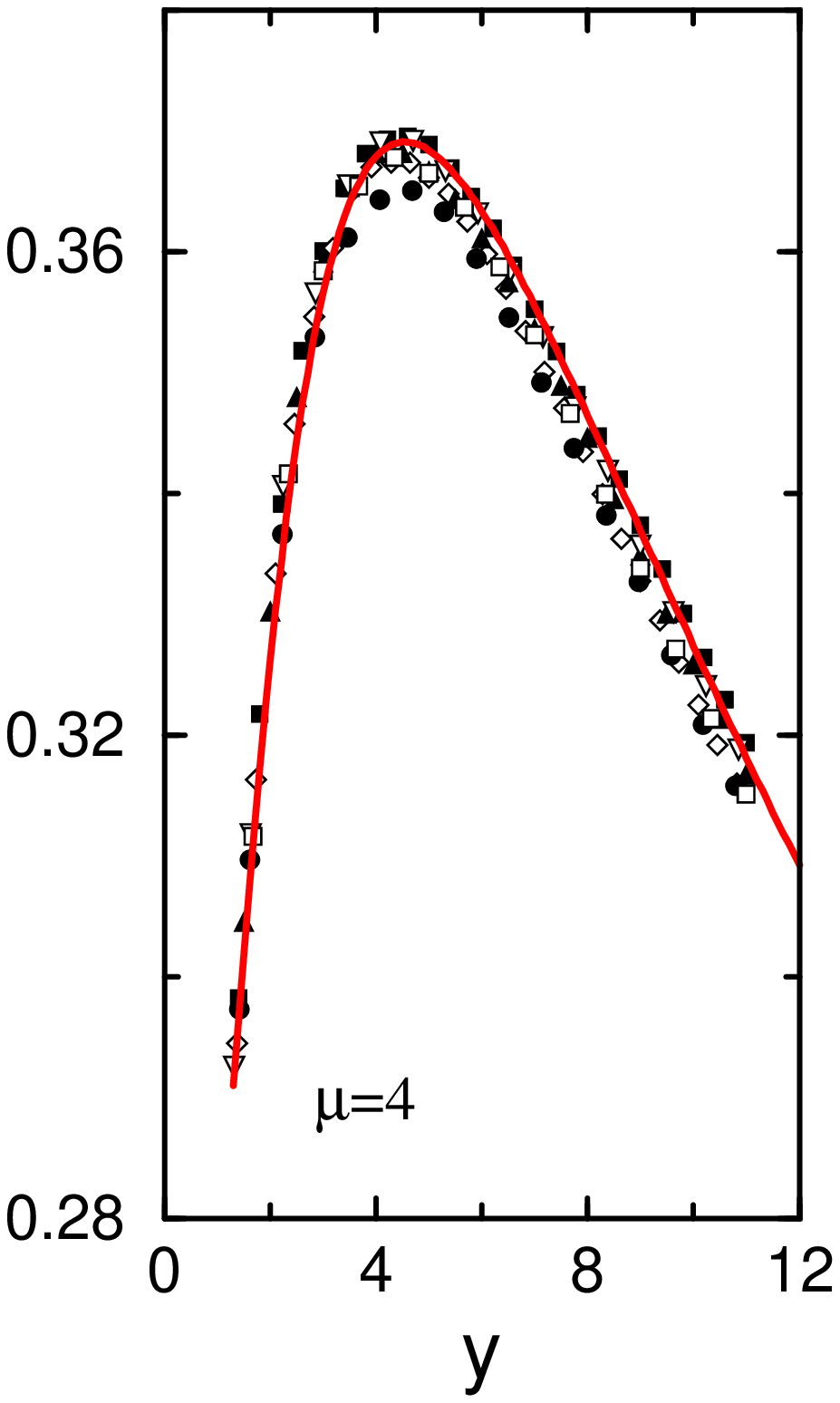}
\caption[]{Integrated space-time response of the $2D$ Glauber-Ising model. 
After {\protect \cite{Henk03b}}.}
\label{Abb2}
\end{figure}

After these preparations, we can now present numerical Monte Carlo (MC) data
and compare them with the predictions (\ref{2:autoR},\ref{2:Rtsr}). We consider
the thermoremanent magnetization $\rho(t,s)$ and subtract off the leading
finite-time correction according to (\ref{3:rho}). For $T<T_c$ this leads to
the parameter values collected in table~\ref{tab2}, see \cite{Henk03b} for
details. Then the MC data both {\em at} $T=T_c$ and for $T<T_c$ for $\rho(t,s)$
are in full agreement with (\ref{3:rho}) \cite{Henk01,Henk03b}, in both $2D$ 
and $3D$. Here we present a direct test of Galiliei-invariance by considering 
the space-time integrated response
\begin{equation}
\frac{\D\rho(t,s;\mu)}{\D\Omega} = T \int_0^s\!\D u\int_0^{\sqrt{\mu s\,}}
\!\D r\, r^{d-1} R(t,u;\vec{r}) = r_0 s^{d/2-a}\rho^{(2)}(t/s,\mu)
\end{equation}
with an explicitly known expression for $\rho^{(2)}$ following from 
(\ref{2:Rtsr}) and the leading finite-time
correction is already subtracted off \cite{Henk03b}. Since all non-universal
parameters were determined before and are listed in table~\ref{tab2}, this 
comparison between simulation and local scale-invariance is parameter-free.
The result in shown in figure~\ref{Abb2} in $2D$ and we find a perfect 
agreement. A similar results holds in $3D$ \cite{Henk03b}. 

This direct evidence in favour of Galilei-invariance\index{Galilei-invariance} 
in the phase-ordering kinetics of the Glauber-Ising\index{Glauber-Ising model} 
model is all the more remarkable since the zero-temperature time-dependent 
Ginzburg-Landau equation (TDGL), which is usually thought to describe the same 
system (e.g. \cite{Bray94}), does not have this symmetry. Indeed 
a recent second-order result for $R(t,s)$ 
does not agree with (\ref{2:autoR}) \cite{Maze04} (similar corrections
also arise at $T=T_c$ \cite{Cala03}). On the other hand, 
$\lambda_C=1$ from the exact solution of the $1D$ Glauber-Ising model at $T=0$ 
\cite{Godr02}, while $\lambda_C\simeq 0.6006\ldots$ in the 
$1D$ TDGL \cite{Bray95}, implying that these two models belong to distinct
universality classes. 

Confirmations of (\ref{2:autoR},\ref{2:Rtsr}) 
in exactly solvable models are reviewed in \cite{Henk02}. 

\section{Influence of noise}
%
We now wish to review the present state of theoretical arguments 
\cite{Henk03,Pico04} in order
to understand from where the recent numerical evidence in favour of a 
larger dynamical symmetry
than mere scale-invariance in ageing phenomena might come from. 
We shall do this here for phase-ordering kinetics. Then $z=2$ and we have to
consider the Schr\"odinger group and Schr\"odinger-invariant systems. 
For simplicity, we often set $d=1$. From the following discussion, the
importance of  Galilei-invariance will become clear, see
also (\ref{2:GalInv}). 

{\bf A)} Consider the free Schr\"odinger equation 
$(2{\cal M}\partial_t - \partial_r^2)\phi=0$ 
where ${\cal M}={\rm i} m$ is fixed. While an element of the Schr\"odinger 
group acts projectively (i.e. up to a known companion function \cite{Nied72})
on the wave function $\phi$, we can go over to a true representation 
by treating $\cal M$ as an additional variable. 
Following \cite{Giul96}, we define a new coordinate $\zeta$ and a 
new wave function $\psi$ by
\begin{equation} \label{Henkel:gl3-17}
\phi(t,r) = \frac{1}{\sqrt{2\pi}} \int_{\mathbb{R}} \!{\rm d}\zeta\, 
e^{-{\rm i}{\cal M}\zeta} \psi(\zeta,t,r)
\end{equation}
We denote time $t$ as the zeroth coordinate and 
$\zeta$ as coordinate number $-1$.

We inquire about the maximal kinematic group in this case 
\cite{Henk03}. Now,  
the projective phase factors can be
absorbed into certain translations of the variable $\zeta$ \cite{Henk03}.
Furthermore, the free Schr\"odinger equation becomes
\begin{equation} \label{Henkel:gl3-18}
\left( 2{\rm i} \partial_{\zeta} \partial_t + 
\partial_r^2 \right) \psi(\zeta,t,r) = 0
\end{equation}
In order to find the maximal kinematic symmetry of this equation, 
we recall that the three-dimensional Klein-Gordon equation 
$\sum_{\mu=-1}^{1}\partial_{\mu}\partial^{\mu} \Psi(\vec{\xi})=0$ has the
$3D$ conformal algebra $\mathfrak{conf}_3 \cong so(4,1)$ as maximal
kinematic symmetry. By making the following change of variables
\begin{equation}
\zeta = \left( \xi_0 + {\rm i} \xi_{-1} \right)/2 \;\; , \;\;
t = \left( -\xi_0 + {\rm i} \xi_{-1} \right)/2 \;\; , \;\;
r = \xi_1 \sqrt{{\rm i}/2} 
\end{equation}
and setting $\psi(\zeta,t,r)=\Psi(\vec{\xi})$, the $3D$ Klein-Gordon equation
reduces to (\ref{Henkel:gl3-18}). Therefore, for variable masses $\cal M$, 
the maximal kinematic symmetry algebra of the 
free Schr\"odinger equation in $d$ dimensions is isomorphic to the conformal 
algebra $\mathfrak{conf}_{d+2}$ and we have the inclusion of the complexified
Lie algebras\index{Schr\"odinger group} \index{conformal group}
$(\mathfrak{sch}_d)_{\mathbb{C}} \subset (\mathfrak{conf}_{d+2})_{\mathbb{C}}$
\cite{Henk03,Burd73}. 

{\bf B)} The Galilei-invariance of the free Schr\"odinger equation requires the
existence of a formal `complex conjugate' $\phi^*$ of the order parameter 
$\phi$. On the other hand, a common starting point in the description of 
ageing phenomena is a Langevin equation which may be turned into a field
theory using the Martin-Siggia-Rose (MSR) formalism \cite{Jans92,Card96} 
and which involves besides $\phi$ the response field $\wit{\phi}$. 
If we identify $\phi^*=\wit{\phi}$ and use (\ref{Henkel:gl3-17}) 
together with the assumption that $\psi$ is real to define the complex 
conjugate, then the causality\index{causality} condition that
\begin{equation}
R(t,s;\vec{r}) = \langle \phi(t,\vec{r})\wit{\phi}(s,\vec{0})\rangle = 
\langle \phi(t,\vec{r})\phi^*(s,\vec{0})\rangle
\end{equation}
vanishes for $t<s$, follows naturally (and similarly for 
three-point response functions) \cite{Henk03}. Therefore, the calculation of
response and of correlation functions from a dynamical symmetry 
should be done in the same way. 

{\bf C)} So far, we have concentrated exclusively in applications of local 
scale-invariance to finding the form of response functions while the 
determination of correlation functions was not yet adressed. We shall do so
now and consider the Langevin equation (with $D^{-1}=2{\cal M}$) \cite{Pico04}
\begin{equation} \label{4:Lang}
\partial_t \phi = -D \frac{\delta {\cal H}}{\delta \phi} -D v(t) \phi +\eta
\end{equation}
where $\cal H$ is the usual Ginzburg-Landau functional, $v(t)$ is a
time-dependent Lagrange multiplier which will be chosen to produce the
constraint $C(t,t)=1$ and $\eta$ is an uncorrelated gaussian noise describing 
the coupling to a heat bath such that $\langle\eta\rangle=0$ and 
$\langle\eta(t,\vec{r})\eta(t',\vec{r}')\rangle=2DT \delta(t-t')
\delta(\vec{r}-\vec{r}')$. Another source of noise comes from the initial
conditions and we shall always use an uncorrelated initial state such that
\begin{equation} \label{4:aini}
C(0,0;\vec{r}) = \langle\phi(0,\vec{r})\phi(0,\vec{0})\rangle 
= a_0 \delta(\vec{r})
\end{equation}
where $a_0$ is a constant. 

The MSR action of (\ref{4:Lang}) reads 
$S[\phi,\wit{\phi}]=S_0[\phi,\wit{\phi}]+S_b[\phi,\wit{\phi}]$
where
\begin{subeqnarray}
S_0[\phi,\wit{\phi}] &=& \int\!\D t \D\vec{r}\: 
\wit{\phi} \left( \frac{\partial\phi}{\partial t} 
+D\frac{\delta{\cal H}}{\delta \phi} + Dv(t)\phi\right) \\
S_b[\phi,\wit{\phi}] &=& -DT\int\!\D t \D\vec{r}\: \wit{\phi}(t,\vec{r})^2 -
\frac{a_0}{2} \int\!\D\vec{r}\: \wit{\phi}(0,\vec{r})^2 
\end{subeqnarray}
and we used (\ref{4:aini}), see \cite{Maze04}. Here $S_0$ describes the 
noiseless part of the action while the thermal and the initial noise are
contained in $S_b$. Finally, the potential $v(t)$ can be absorbed into a
gauge transformation; for example if $\phi^{(0)}$ is a solution of the free
Schr\"odinger equation, then $\phi=\phi^{(0)} k(t)$ solves the Schr\"odinger
equation with the potential $v(t)$ and where
\begin{equation}
k(t) := \exp\left( -D \int^{t} \!\D u\: v(u) \right)
\end{equation}
The realization of the Schr\"odinger algebra with $v(t)\ne 0$ is easily
found \cite{Pico04}. 

We now assume in addition to dynamical scaling that $\cal H$ is such that 
{\it at temperature $T=0$, the theory is\index{Galilei-invariance}
Galilei-invariant} \cite{Pico04}. This looks physically reasonable and we 
now explore some
consequences of this hypothesis. We denote by $\langle\cdot\rangle_0$ an 
average carried out using only the noiseless part
$S_0$ of the action. The Bargman 
super\-se\-lec\-tion rules state that 
$\langle \underbrace{\phi\cdots\phi}_n
\underbrace{\wit{\phi}\cdots\wit{\phi}}_m\rangle_0 =0$ if $n\ne m$.  
First, the response function\index{response function} is
\begin{eqnarray}
R(t,s;\vec{r}) &=& 
\left\langle\phi(t,\vec{r})\wit{\phi}(s,\vec{0})\right\rangle 
= \left\langle\phi(t,\vec{r})\wit{\phi}(s,\vec{0})
\exp\left(-S_b[\phi,\wit{\phi}]\right)\right\rangle_0 
\nonumber \\
&=& \left\langle\phi(t,\vec{r})\wit{\phi}(s,\vec{0})\right\rangle_0 =: 
R_0(t,s;\vec{r}) \label{4:R}
\end{eqnarray}
where in the last line the exponential was expanded and the 
Bargman super\-se\-lec\-tion rule was used. 
Here $R_0$ is the noiseless response and 
the form (\ref{2:autoR},\ref{2:Rtsr}) of Schr\"odinger-invariance 
is recovered if $v(t)=(2{\cal M})(1+a-\lambda_R/2) t^{-1}$ \cite{Pico04}. 
In other words, under the stated hypothesis, the response function is
independent of the noises. This is certainly in agreement with the explicit
model calculations reviewed in section~3.  

Second, we now obtain the autocorrelation\index{correlation function} 
function. As before \cite{Pico04} 
\begin{eqnarray}
C(t,s) &=& \left\langle\phi(t,\vec{r})\phi(s,\vec{r})
\exp\left(-S_b[\phi,\wit{\phi}]\right)\right\rangle_0 
\nonumber \\
&=& DT\int\!\D u\D\vec{R}\, R_0^{(3)}(t,s,u;\vec{R}) +
\frac{a_0}{2} \int\!\D\vec{R}\, R_0^{(3)}(t,s,0;\vec{R})
\end{eqnarray}
where $R_0^{(3)}(t,s,u;\vec{R}) =\langle\phi(t,\vec{y})\phi(s,\vec{y})
\wit{\phi}(u,\vec{R}+\vec{y})^2\rangle_0$. In contrast with the response
function, the autocorrelation function contains {\em only} noisy terms and in
fact vanishes in the absence of noise. By hypothesis, Schr\"odinger-invariance
holds\index{Schr\"odinger-invariance} 
for the noiseless theory and the\index{three-point function} 
three-point function $R_0^{(3)}$ is
fixed up to a scaling function of a single variable \cite{Henk94,Pico04}. 

Working out the asymptotic behaviour of $C(t,s)$ for $y=t/s\to\infty$ 
according to (\ref{1:C}) and
comparing with the response function (\ref{4:R}), we find that
{\it for any coarsening system with a disordered initial state (\ref{4:aini}) 
and whose noiseless part is Schr\"odinger-invariant, the
\index{autocorrelation exponent}\index{autoresponse exponent} relation 
$\lambda_C=\lambda_R$ holds true} \cite{Pico04}. For the first time
a general sufficient criterion for this exponent relation is found. 

The autocorrelator scaling function becomes for phase-ordering 
\begin{equation}
f_C(y) = \frac{a_0}{2} y^{\lambda_C/2}(y-1)^{-\lambda_C} \Phi\left(
\frac{y+1}{y-1}\right)
\end{equation}
but the scaling function $\Phi(w)$ is left undetermined by 
Schr\"odinger-invariance. 
If in addition we require that $C(t,s)$ should be non-singular at $t=s$,
the asymptotic behaviour $\Phi(w)\sim w^{-\lambda_C}$ for $w\to\infty$ follows.
Provided that form should hold true for all values of $w$, we would obtain
approximately 
\begin{equation} \label{4:approx}
f_C(y) \approx f_0 \left(\frac{(y+1)^2}{4y}\right)^{-\lambda_C/2}
\end{equation}
Indeed, this is found to be satisfied for several ageing spin systems with
an underlying free-field theory \cite{Pico04}. 
On the other hand, (\ref{4:approx}) does not hold true in the Glauber-Ising
model. Work is presently being carried out in order to describe $f_C(y)$ in 
this model and will be reported elsewhere \cite{Henk04}.

It is a pleasure to thank M. Pleimling, A. Picone and J. Unterberger for the
fruitful collaborations which led to the results reviewed here. This work
was supported by CINES Montpellier (projet pmn2095) and by the 
Bayerisch-Franz\"osisches Hochschulzentrum (BFHZ).

%


\begin{thebibliography}{8.}
\addcontentsline{toc}{section}{References}

\bibitem{Stru78} L.C.E. Struik: {\it Physical ageing in amorphous polymers and
other materials} (Elsevier, Amsterdam 1978).
\bibitem{Bray94} A.J. Bray: Adv. Phys. {\bf 43}, 357 (1994). 
\bibitem{Bouc00} J.P. Bouchaud in M.E. Cates, M.R. Evans (eds)
{\it Soft and fragile matter} (IOP Press, Bristol 2000).
\bibitem{Godr02} C. Godr\`eche, J.M. Luck: J. Phys. Cond. Matt. {\bf 14},
1589 (2002).
\bibitem{Cugl02} L.F. Cugliandolo: in {\it Slow Relaxation and
non equilibrium dynamics in condensed matter}, Les Houches Session 77 July 2002,
J-L Barrat, J Dalibard, J Kurchan, M V Feigel'man eds 
(Springer, Heidelberg 2003)
\bibitem{Fish88} D.S. Fisher, D.A. Huse: Phys. Rev. {\bf B38}, 373 (1988).
\bibitem{Pico02} A. Picone, M. Henkel: J. Phys. {\bf A35}, 5575 (2002).
\bibitem{Jans89} H.K. Janssen, B. Schaub, B. Schmittmann:
Z. Phys. {\bf B73}, 539 (1989).
\bibitem{Sche03} G. Schehr, P. Le Doussal: Phys. Rev. {\bf E68}, 046101
(2003). 
\bibitem{Yeun96} C. Yeung, M. Rao, R.C. Desai: Phys. Rev. {\bf E53},
3073 (1996).
\bibitem{Henk02a} M. Henkel, M. Pae{\ss}ens, M. Pleimling: 
Europhys. Lett. {\bf 62}, 644 (2003)
\bibitem{Bert99} L. Berthier, J.L. Barrat, and J. Kurchan: Eur. Phys. J. 
{\bf B11}, 635 (1999).
\bibitem{Henk03e} M. Henkel, M. Pae{\ss}ens, M. Pleimling:
{\tt cond-mat/0310761}.
\bibitem{Henk94} M. Henkel: J. Stat. Phys. {\bf 75}, 1023 (1994). 
\bibitem{Henk97} M. Henkel: Phys. Rev. Lett. {\bf 78}, 1940 (1997).
\bibitem{Henk02} M. Henkel: Nucl. Phys. {\bf B641}, 405 (2002).
\bibitem{Card96} J.L. Cardy: {\it Scaling and Renormalization in 
Statistical Mechanics} (Cambridge University Press, Cambridge 1996).
\bibitem{Bela84} A.A. Belavin, A.M. Polyakov, A.B. Zamolodchikov:
Nucl. Phys. {\bf B241}, 333 (1984).
\bibitem{Henk99} M. Henkel: {\it Phase Transitions and Conformal Invariance}
(Springer, Heidelberg 1999).
\bibitem{Nied72} U. Niederer: Helv. Phys. Acta {\bf 45}, 802 (1972).
\bibitem{Hage72} C.R. Hagen: Phys. Rev. {\bf D5}, 377 (1972).
\bibitem{ORai01} L. O'Raifeartaigh and V.V. Sreedhar: Ann. of Phys. 
{\bf 293}, 215 (2001). 
\bibitem{Henk03} M. Henkel, J. Unterberger: Nucl. Phys. {\bf B660}, 
407 (2003).
\bibitem{Plei01} M. Pleimling, M. Henkel: Phys. Rev. Lett. {\bf 87}, 
125702 (2001). 
\bibitem{Cher04} R. Cherniha, M. Henkel: {\tt math-ph/0402059}.
\bibitem{Jans92} H.K. Janssen: in G. Gy\"orgyi et {\em al.} (eds) {\it
From Phase transitions to Chaos}, World Scientific (Singapour 1992), p. 68
\bibitem{Henk01} M. Henkel, M. Pleimling, C. Godr\`eche, J.-M. Luck:
Phys. Rev. Lett. {\bf 87}, 265701 (2001).
\bibitem{Barr98} A. Barrat: Phys. Rev. {\bf E57}, 3629 (1998).
\bibitem{Abra89} D.B. Abraham and P.J. Upton, Phys. Rev. {\bf B39}, 736 (1989).
\bibitem{Corb03} F. Corberi, E. Lippiello, M. Zannetti: 
Phys. Rev. {\bf E68}, 046131 (2003).
\bibitem{Henk03b} M. Henkel, M. Pleimling: Phys. Rev. {\bf E68}, 
065101(R) (2003). 
\bibitem{Maze04} G.F. Mazenko: Phys. Rev. {\bf E69}, 016114 (2004).
\bibitem{Cala03} P. Calabrese, A. Gambassi: Phys. Rev. {\bf E67}, 
036111 (2003). 
\bibitem{Bray95} A.J. Bray, B. Derrida: Phys. Rev. {\bf E51}, R1633 (1995).
\bibitem{Pico04} A. Picone, M. Henkel: {\tt cond-mat/0402196}. 
\bibitem{Giul96} D. Giulini: Ann. of Phys. {\bf 249}, 222 (1996).
\bibitem{Burd73} G. Burdet, M. Perrin, P. Sorba: Comm. Math. Phys. {\bf 34},
85 (1973). 
\bibitem{Henk04} M Henkel, A. Picone, M. Pleimling: to be published. 


\end{thebibliography}
\end{document}